\documentclass[11pt,twoside]{article}  
\usepackage{asp2006}
\usepackage{adassconf}

\begin{document}   

%
%

\paperID{P4.1}

%

\title{A pseudo-parallel Python environment for database curation}

%
%
%
%
%

\markboth{Eckhard Sutorius et al.}{A pseudo-parallel Python environment for database curation}

%
%
%
%

\author{Eckhard Sutorius, Johann Bryant, Ross Collins, Nicholas Cross, Nigel Hambly, Mike Read}
\affil{Scottish Universities Physics Alliance (SUPA), Institute for Astronomy, School of Physics, University of Edinburgh, UK}

%

\contact{Eckhard Sutorius}
\email{etws@roe.ac.uk}

%
%
%

\paindex{Sutorius, E.~T.}
\aindex{Bryant, J.}     
\aindex{Collins, R.~S.}
\aindex{Cross, N.~J.}
\aindex{Hambly, N.~C.}
\aindex{Read, M.~A.}

%



\begin{abstract}          
One of the major challenges providing large databases like the WFCAM Science
Archive (WSA) is to minimize ingest times for pixel/image metadata
and catalogue data. In this article we describe how the pipeline processed
data are ingested into the database as the first stage in building a release
database which will be succeeded by advanced processing (source merging,
seaming, detection quality flagging etc.). To accomplish the ingestion
procedure as fast as possible we use a mixed Python/C++ environment and run
the required tasks in a simple parallel modus operandi where the data are
split into daily chunks and then processed on different computers. The
created data files can be ingested into the database immediately as they
are available. This flexible way of handling the data allows the most usage
of the available CPUs as the comparison with sequential processing shows.
\end{abstract}

%
%

\section{Introduction}
The WFCAM Science Archive (\htmladdnormallinkfoot{WSA}{http://surveys.roe.ac.uk/wsa/}; Hambly et al. 2007, Collins et al. 2006) holds the image and catalogue 
data products generated by the Wide Field Camera (WFCAM) on UKIRT (United 
Kingdom Infrared Telescope). The data comprise pipeline processed 
multi-extension FITS files (multiframes) containing pixel/image and 
catalogue data for four detectors at one pointing. The latter contains all 
detections of stacked multiframes. The data are pipeline processed at the 
Cambridge Astronomical Survey Unit (CASU) and transferred to Edinburgh where 
the Wide Field Astronomy Unit (WFAU) processes it for ingestion into the 
database. 
Since the release database contains advanced products which can take a lot
of CPU time to produce, it is preferable to carry out the ingest procedure as 
fast as possible. Another aspect is that the pixel/image and catalogue data 
need to be ingested completely before further processing is done. Another
constraint is the uniqueness of multiframes and detections. Each multiframe 
has to have a unique identifier across the whole database and each detection
must be unique for a given survey.

\section{The Curation Usecases}
To ingest the large amount of data into the WSA as the first stage in 
building a release database, a set of Curation Usecases (CUs) have been 
designed. They are coded in a Python/C++ environment, where C++ is used where 
high performance is needed and Python to facilitate an easy to use 
object-oriented environment. Table~\ref{tab:P4.1_tab1} shows an overview of
the ingest CUs and the average volume of data to be processed per observing 
night.

\begin{deluxetable}{lll}%
\tablecaption{The ingest curation usecases\label{tab:P4.1_tab1}}
\tablehead{
\colhead{CU} & \colhead{task} & \colhead{volume/observing night}}
\startdata
1 & Data transfer from CASU to WFAU & $\sim 80$~GByte\\[5pt]
2 & Creation of compressed images (JPEGs) & $\sim 12000$~JPEGs\\[5pt]
3 & \parbox{6cm}{Extraction, process and ingest of multiframe metadata} & $\sim 3000$~files \\[10pt]
4 & \parbox{6cm}{Extraction, process and ingest of catalogue data} &  $\sim 3\cdot10^6$~detections \\[10pt]
\enddata
\end{deluxetable}

In the following we will concentrate on CUs 2 to 4, as the data transfer (CU1) 
is described in detail in Bryant et al. (2008). After transfer the data are 
split into daily chunks and then processed on multiple computers. \\

The following dependencies need to be observed to avoid duplicate entries 
or missing data. Each FITS file gets assigned a unique multiframe ID 
associating data across the database with its source. Also each object in a 
catalogue gets assigned a unique object ID associating data across the database
with its original detection. To update the database with the paths to 
compressed images (CU2), the general FITS file metadata has to be ingested 
beforehand (CU3). And finally catalogue data can only be processed (CU4) if the
corresponding image metadata is available (CU3). 

\section{Parallelization}
\begin{figure}[h]
\plotone{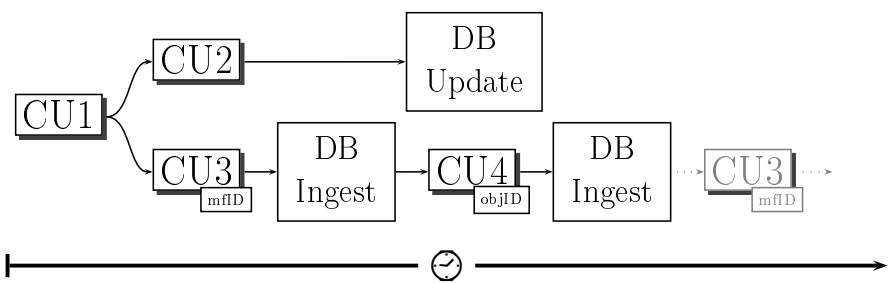}
\caption{Non-parallel CU execution. The creation of the unique identifiers is indicated by the boxes denoted mfID (multiframe ID) and objID (object ID), respectively.}
\label{fig:P4.1_fig1}
\end{figure}

The normal procedure of executing the CUs for small amounts of data would 
comprise of a run of each CU followed by an ingest as shown in 
Figure~\ref{fig:P4.1_fig1}. This might then be followed again by the whole
procedure for the next batch of data. As the creation of compressed images 
(CU2) itself can be done independently, it can be decoupled from image and 
catalogue data processing. Since this procedure is linear, unique IDs for 
pixel files and catalogue detections are applied during CU3 and CU4, 
respectively.\\

\begin{figure}[h]
\plotone{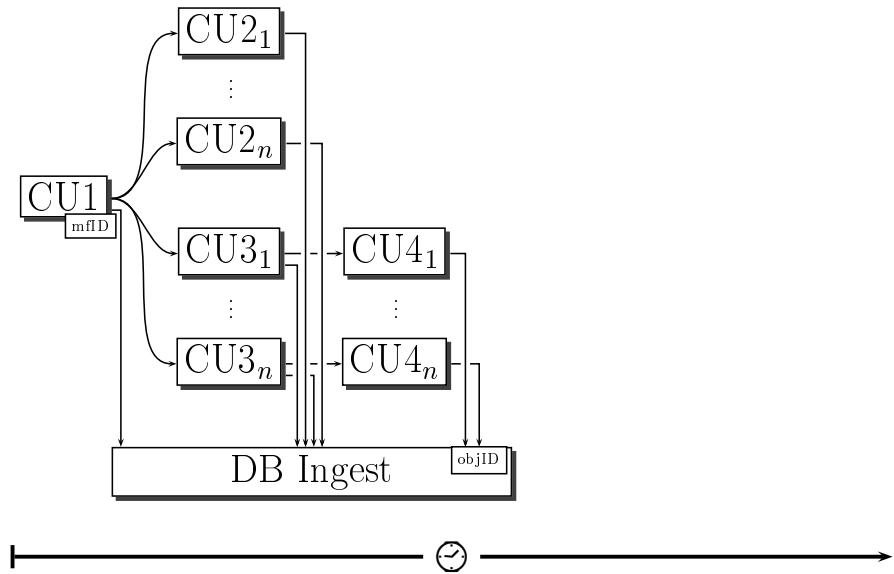}
\caption{Parallelized CU execution. The creation of the unique identifiers is indicated by the boxes denoted mfID (multiframe ID) and objID (object ID), respectively.}
\label{fig:P4.1_fig2}
\end{figure}

To improve CPU usage and speed up the process for a whole cycle the following 
enhancements were applied:
\begin{enumerate}
\item Files get multiframe IDs depending on the largest mulitframe ID in the 
database directly after transfer (CU1). The database is accordingly updated.
\item The CU4 object ID is turned into a temporary negative, only per day 
unique ID. This way we avoid overlaps with already in the database existing 
(positive) object IDs. It can be translated into a global unique ID directly 
after ingest with a simple addition to the last maximal object ID. 
\item The ingest process is completely de-coupled from extraction and 
processing.
\end{enumerate}
The first two steps allow us to process metadata simultaneously on different 
computers. The last one allows the ingest to be run on available ingestable 
data while new data is processed. Since ingest can take as much time as 
processing, this can nearly halve the run time of the full cycle. 
Figure~\ref{fig:P4.1_fig2} shows the flow chart for the final design.

\section{Automation}
A data daemon has been created that checks the available computers, their load,
and the tasks that need to be executed as well as tasks already running. 
At the moment it suggests the distribution of these tasks to enhance usage 
of CPUs on the pixel servers. The database operator
then takes the final decision in running the tasks. In addition to the data 
daemon the ingester daemon can be run, checking automatically for ingestable 
data and ingesting them into the database. This maximises CPU usage on the 
database server.\\

Since these daemons can be run automatically, we are investigating the best 
way to run individual tasks remotely from a master computer. A number of 
Python-related solutions exist and are described in the \htmladdnormallinkfoot{Python Parallel Processing Wiki}{http://wiki.python.org/moin/ParallelProcessing}.

\section{Results}
Running each of the CUs 2 to 4  for 30 nights of data on five computers at the 
same time and ingesting the data as soon as it was available improved the 
total run times as follows:
\begin{itemize}
\item CU2: up to 4$\times$ faster;
\item CU3: up to 3$\times$ faster;
\item CU4: up to 2$\times$ faster.
\end{itemize}
The differences are due to the different ratios of time needed to process 
and ingest the data. During normal operations the gain is about 30-40\% for 
concurrent runs of all CUs.

\end{document}